\documentclass[a4paper]{article}

\usepackage{icrc2013}
\usepackage[english]{babel}
 \pdfoutput=1 
 
\title{Study of high energy cosmic ray acceleration in Tycho SNR with VERITAS}

\shorttitle{Study of high energy cosmic ray acceleration in Tycho SNR with VERITAS}

\authors{
Nahee Park$^{1}$
for the VERITAS Collaboration.
}

\afiliations{
$^1$ Enrico Fermi Institute, University of Chicago \\
}

\email{nahee@uchicago.edu}

\abstract{Supernova Remnants (SNRs) are broadly accepted as the main accelerators of Galactic cosmic rays (GCRs) with energies up to the knee region. Recent measurements of pion bumps in IC 443 and W 44 by $Fermi$-LAT show indirect evidence of the acceleration of hadronic particles in SNRs. But, whether SNRs are the powerhouses for GCR acceleration all the way up to the knee region still remains an unsolved question. Tycho is a promising target for this study because it has been widely studied in multi-wavelength observations from IR to TeV and it is a young type Ia SNR located in a relatively clean environment. Though recently developed models generally agree on the likely hadronic origin of the gamma-ray emission from Tycho, the details of the models vary considerably because the current data in the GeV-TeV range are weakly constraining. Since the initial detection, VERITAS has increased its data size by more than 40\%. We also recently upgraded the telescope cameras and the analysis packages, which will allow us to extend the measurements toward both lower and higher energies. In this talk, I present updates on the VHE gamma-ray measurements of Tycho with VERITAS, focusing on the interpretation of the additional data in the context of hadronic GCR acceleration models.} 


\begin{document}
\maketitle

\section{Introduction}

Galactic cosmic rays (GCRs) observed at Earth are composed of $>$ 98 \% protons and heavier nuclei and $<$ 2 \% electrons and positrons \cite{bib:simpson83}. Since the discovery of CRs, the acceleration mechanisms and origins of CRs have been intensively studied. Ginzburg and Syravotskii \cite{bib:ginzburg64} suggested that supernova remnants (SNRs) are the sources of GCRs, which has become the generally accepted hypothesis. A modest efficiency of $\sim10$\% in converting the kinetic energy of supernova shocks into particle acceleration and the frequency of SNR in our Galaxy can explain the observed flux of cosmic rays. Also, diffusive shock acceleration (DSA) theories suggest that SNRs can accelerate particles up to $10^{15}$ eV \cite{bib:bell78, bib:hillas05}. DSA predicts the intrinsic spectral index of accelerated cosmic rays to be $\sim$ -2.0, which is in agreement with the measured index of CRs after propagation corrections. 

Indirect evidence for the acceleration of the leptonic component of cosmic rays up to the TeV energy range has been found in X-ray observations of SNRs (e.g., SN 1006 \cite{bib:koyama95}). Non-thermal X-ray emissions from the rims of SNRs suggest the existence of high-energy electrons with energies $\geq$ 200 TeV, under the assumption of a moderate magnetic field strength of $\sim$ 10 $\mu$G. Observing high-energy gamma rays from SNRs was expected to provide indirect evidence of the acceleration of high energy hadronic components via pion decay from hadronic interactions of particles with the surrounding interstellar medium (ISM). In fact, several SNRs were detected in the GeV and TeV energy ranges by space-borne experiments such as $Fermi$-LAT and ground-array telescopes such as VERITAS. However, studying hadronic acceleration in SNRs with gamma rays is not straightforward because the inverse Compton scattering of leptonic components with background radiation can also generate high-energy gamma rays. Detailed modeling of the evolution of SNRs in their environments is required to estimate the contribution of hadronic and leptonic origins of gamma rays. This is a challenging task because many of the relevant quantities are unknown. 

Recently $Fermi$-LAT reported the detection of the pion-decay signature from W 44 and IC 443 \cite{bib:ackermann13}. Both SNRs are very bright in the GeV energy band and known for interactions with nearby dense materials. With four years of data and improvements in the analysis method, $Fermi$-LAT could resolve the pion bump at $\sim$ 200 MeV in both SNRs, showing the existence of hadronic components with energies higher than several GeV. Although it is possible to explain this feature with a leptonic scenario, this requires an additional abrupt break in the electron spectrum. These measurements provide strong evidence for the acceleration of protons in the remnants. However, whether SNRs are the powerhouses for GCRs all the way up to the $10^{15}$ eV still remains an unsolved question. Also, the detailed acceleration mechanism including the efficiency and energetics requires more study.

\section{Tycho SNR as a GCR accelerator}

Tycho is the remnant of a historic SN that was observed in 1572 \cite{bib:stephane02}. Tycho was categorized as resulting from a Type Ia SN \cite{bib:baade45}, which was confirmed by the light echoes from Tycho's explosion \cite{bib:rest08, bib:krause08}. Expansion measurements of the remnant match with a SNR transitioning into the Sedov phase \cite{bib:reynoso97, bib:kamper78,bib:katsuda10}. Estimated distances to the remnant have varied from 2 kpc to 5 kpc. A recent X-ray study yielded 4.0 $\pm$ 1.0 kpc \cite{bib:hayato10}. As one of the younger SNRs, Tycho has been studied in many wavelengths. Tycho has a shell-type morphology at radio wavelengths with enhanced emission along the northeastern edge of the remnant \cite{bib:dickel91, bib:stroman09}. 
X-ray images show strong non-thermal emission along the rim \cite{bib:katsuda10}. Deep exposures with Chandra revealed strips of non-thermal emission from the interior of the remnant, which was suggested to be evidence of turbulence in magnetic fields that can accelerate particles up to $10^{15}$ eV \cite{bib:eriksen11}. A lack of thermal X-rays at the forward shock\cite{bib:hwang02} and a relatively spherical morphology in radio and X-ray suggest that Tycho may have exploded into a relatively uniform low-density environment. IR measurements with Spitzer showed a possible gradient in the ISM density, denser at the northeast side compared to the southwest side by a factor of 3$\sim$5 \cite{bib:williams13}. Also, the study reported a few regions with significantly higher density located at the northeast and east sides of the remnant. Several studies proposed the interaction of Tycho with a dense molecular cloud (MC) located in the northeast side of the remnant \cite{bib:reynoso99,bib:lee04}, although a more recent study showed that there is no evidence of direct interaction between the SN shock and the MC \cite{bib:tian11}. 

\begin{figure}[t]
 \centering
  \includegraphics[width=0.5\textwidth]{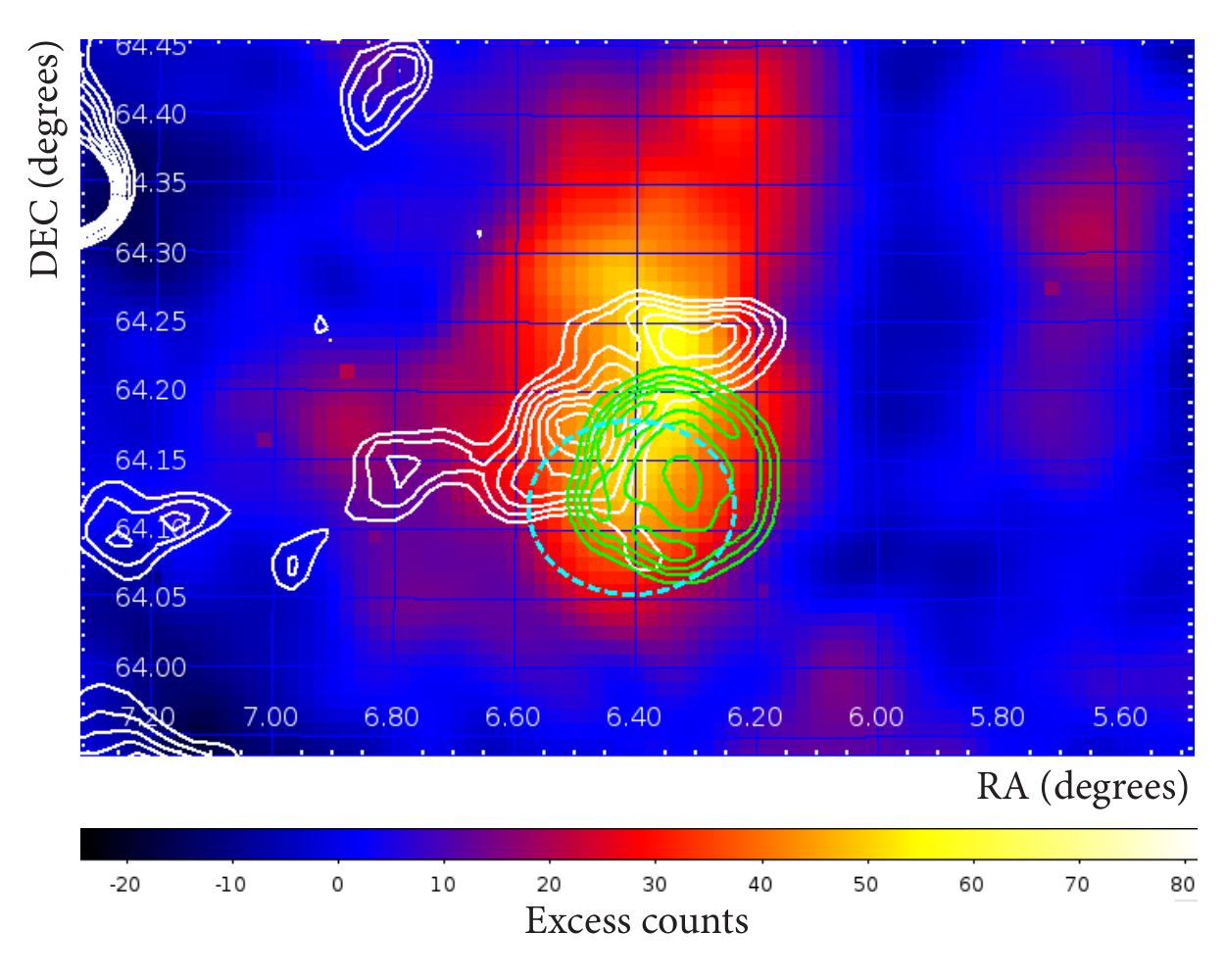}
  \caption{Excess map of gamma-ray events observed by VERITAS in the vicinity of Tycho, collected during the 2008/2009 and 2009/2010 observing seasons~\cite{bib:acciari11}. An squared integration radius of 0.015 $deg^{2}$ was used for the former data set and 0.01 $deg^{2}$ for the latter. The white lines depict $^{12}$CO emission (J = 1-0), integrated over velocity from -68 km/s to -50 km/s, from the high-resolution FCRAO Survey \cite{bib:heyer98}. The 1420 MHz radio map from the Canadian Galactic Plane Survey~\cite{bib:kothes06} appears as green contours. The dashed cyan line indicates the $Fermi$-LAT 95\% error on the centroid of the GeV emission~\cite{bib:giordano12}.}
  \label{fig_tycho_skymap}
 \end{figure}
 
All these studies made Tycho an interesting target and stimulated the observations in high-energy gamma rays. Recently, VERITAS \cite{bib:acciari11} and Fermi \cite{bib:giordano12} discovered high energy gamma-ray emission from Tycho. VERITAS reported a relatively hard spectrum with spectral index of $-1.95\pm0.51_{stat}\pm0.3_{sys}$ covering the 1 $\sim$ 10 TeV energy range, while the spectrum of $Fermi$-LAT covers energies from $\sim$ 300 MeV to 100 GeV with a softer spectral index of $-2.3\pm0.2_{stat}\pm0.1_{sys}$. The TeV sky map can be seen in Figure~\ref{fig_tycho_skymap}. The centroid of the TeV emission from Tycho is slightly displaced toward the northeast side, where a MC was detected, deviating from the center of the remnant by $0.04^{\circ}$. This offset was not statistically significant.

Detailed studies in wide energy range make Tycho a good candidate to study the acceleration of CRs in SNRs. In addition to this, it sits in a relatively clean environment because, as a Type Ia SNR, the progenitor star hasn't modified the environment as in core-collapse SNe. Several models have been proposed to explain the large-scale spectral-energy distribution of Tycho. Although there are on-going efforts to fully account for the hydrodynamical flow pattern, such as studies of accelerations in the reverse shock~\cite{bib:telezhinsky12}, currently available models for Tycho are, in general, based on DSA at the forward shock. While most of the models agree on the hadronic origin of gamma rays, there are still noticeable disagreements in detail. 

Morlino and Caprioli \cite{bib:morlino12} presented a full model of Tycho from radio to X-ray and explained the origin of the high-energy gamma rays as a hadronic component. They concluded that Tycho can accelerate protons up to 500 TeV with $\sim$ 12 \% of the kinetic energy being converted into the acceleration of CRs. While Morlino and Caprioli's model obtained a gamma-ray index of -2.1, which is softer than the predictions from standard DSA and non-linear DSA, Berezhko et al. \cite{bib:berezhko12} supported a gamma-ray index of -2.0. To explain the softer index of GeV emission, Berezhko's model added an extra component originating from the interactions with dense clumps around the SNR.  Zhang et al. \cite{bib:zhang13} suggested that the gamma-ray emission originated from interactions with the nearby MC, which requires the conversion efficiency from kinetic energy into CR acceleration to be at the level of only 1\%. Alternatively, Atoyan and Dermer \cite{bib:atoyan12} explored the possibility for a pure leptonic model to explain the emission by taking into account of the possibility of multiple emission zones. 

\begin{figure}[t]
 \centering
  \includegraphics[width=0.5\textwidth]{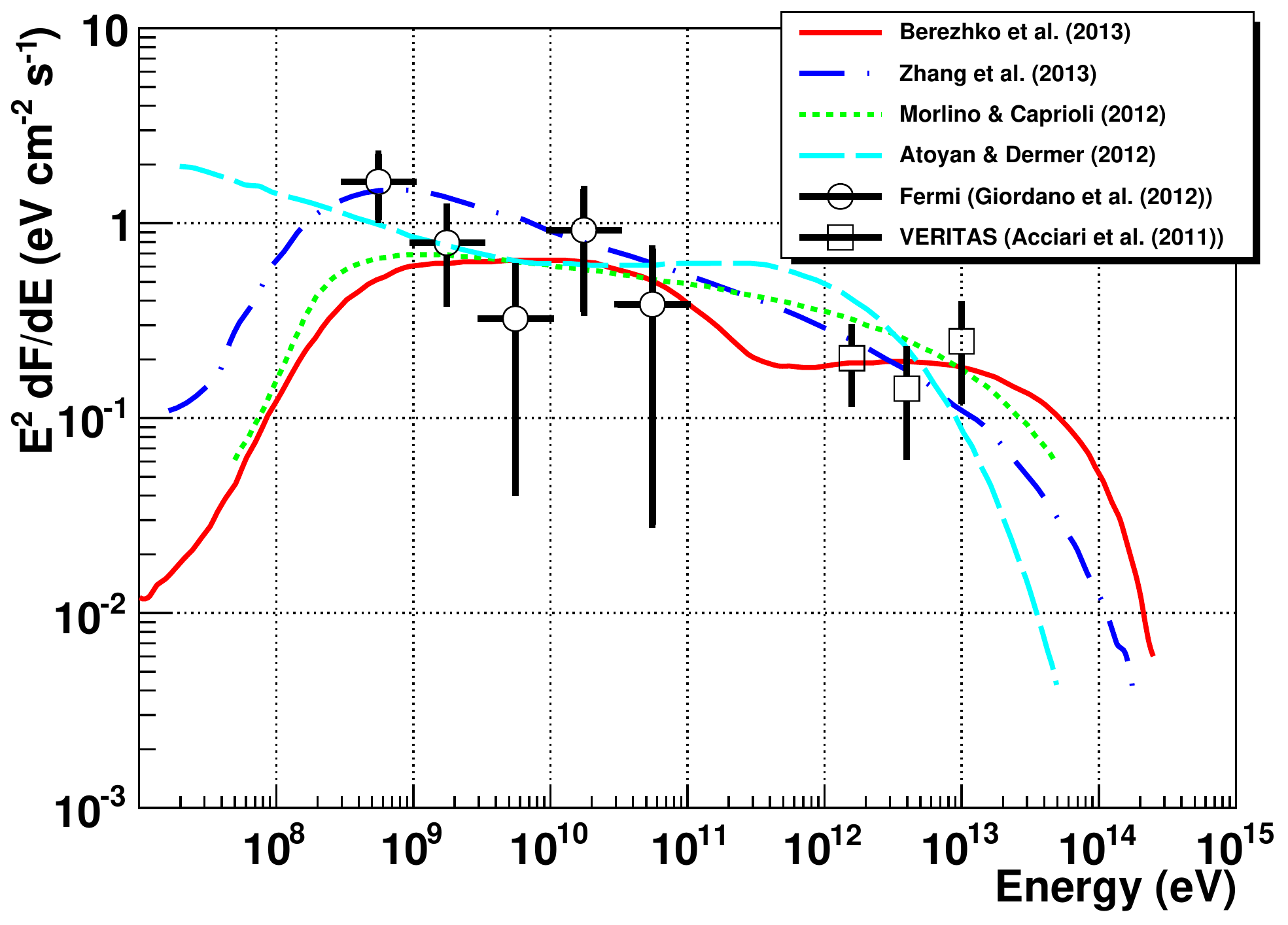}
  \caption{Previously published VERITAS \cite{bib:acciari11} and $Fermi$-LAT data \cite{bib:giordano12} with various models. The red solid line shows the model of Berezhko et al. \cite{bib:berezhko12}, for the case of gamma-ray production in a clumpy ISM. Zhang et al \cite{bib:zhang13}'s model is plotted as blue dotted line, showing the case of emission mainly coming from MC. Morlino \& Caprioli \cite{bib:morlino12}'s model appears as a green dotted line, representing CR production in a homogeneous ISM. The cyan dashed line represents the case of pure leptonic model with multiple emission zones \cite{bib:atoyan12}.}
  \label{fig_tycho_spectrum}
 \end{figure}

Figure~\ref{fig_tycho_spectrum} shows the different model predictions compared to the $Fermi$-LAT and VERITAS measurements. Given the size of the error bars on the currently available data, it is difficult to discriminate between models. Reducing the statistical error bars and extending the spectrum toward both lower and higher energies can provide better constraints. Improvement on the centroid location of high-energy gamma rays will be important for determining the acceleration efficiency and the origin of gamma rays from Tycho.

\section{ VERITAS observations }

VERITAS \cite{bib:holder06, bib:holder08} is an array of four imaging air Cherenkov telescopes located at the base camp of the Fred Lawrence Whipple Observatory in southern Arizona. Each telescope has a camera that covers a total field of view of $3.5^{\circ}$. The configuration of the VERITAS array was changed in summer 2009 by moving one telescope (T1) to make a more symmetric array. The relocation of T1 enhanced the gamma/hadronic discrimination and improved the sensitivity of VERITAS. After the relocation, the array became sensitive to photons with energies between 100 GeV and 30 TeV. VERITAS could detect a point source with 1 \% of the flux of the Crab Nebula at the 5$\sigma$ level with less than 25 hours of exposure with angular resolution of $\sim$ $0.1^{\circ}$ \cite{bib:holder11}. In summer 2012, VERITAS updated its cameras by installing high quantum efficiency PMTs. With the new PMTs, the trigger threshold of VERITAS was reduced, expanding the sensitive energy range down to $\sim$ 85 GeV \cite{bib:kieda13}.

VERITAS has observed Tycho since 2008. Data published in the first discovery paper includes $\sim$ 22 hours of data before the relocation of T1 and $\sim$ 45 hours of data after the relocation. We increased the exposure on the source by $\sim$ 38 hours from January 2010, which includes 19 hours of data with updated cameras. 
 
\section{ Results and discussion }
VERITAS observed the Tycho region for $\sim$ 105 hours over five years. There have been two major upgrades to the telescope configuration of VERITAS in these five years of observations: relocation of one telescope (September 2009) and a camera upgrade (September 2012). To understand the source of emissions, careful systematic studies of the properties of each telescope configuration and the interpretations of combined results are required. Results shown here are part of on-going systematic studies. We used 64 hours of data with the same telescope configuration, which were obtained from 2009 to 2011.

A camera image-fitting algorithm \cite{bib:christiansen12} was used for the analysis. Results of the camera image fitting with a 2D Gaussian function were used for the shower directional reconstruction and the gamma-ray/hadronic separation. The method provides improvements on the angular resolution and the effective area especially for high energy events. Analysis with this 2D Gaussian fitting method can detect a weak source with $\sim$ 20\% less observing time compared to the standard Hillas moment analysis. Results were checked internally with the standard method and with another independent analysis package.

\begin{figure}[t]
 \centering
  \includegraphics[width=0.5\textwidth]{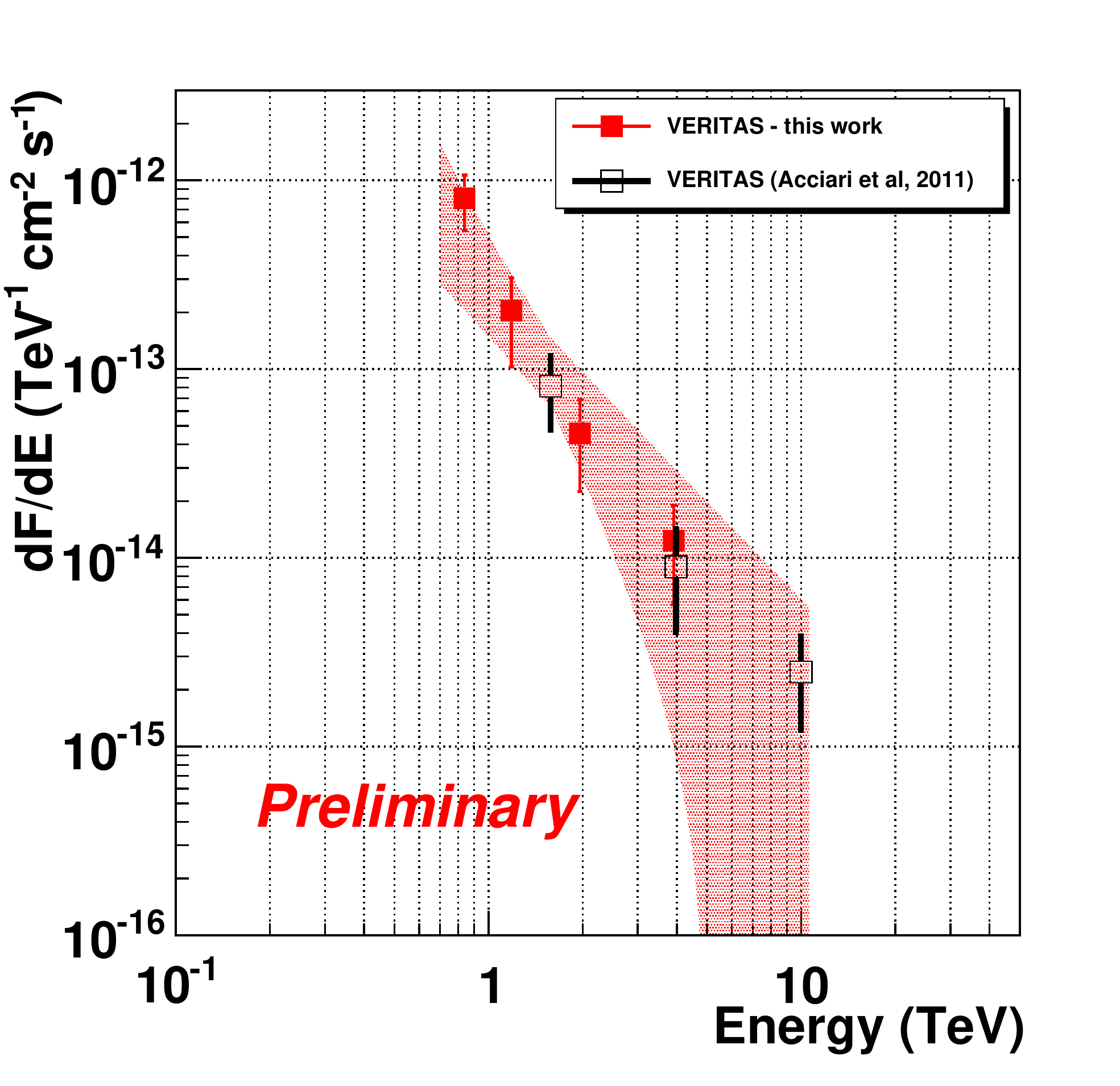}
  \caption{Previously published VERITAS \cite{bib:acciari11} with an updated spectrum. Previous measurements comprise 21.9 hours of data from the 2008/2009 season and 44.7 hours of data from the 2009/2010 season. The spectrum from this work uses $\sim$ 64 hours of data from 2009 $\sim$ 2011. }
  \label{fig_tycho_spectrum_update}
 \end{figure}

An excess of gamma-ray events was found with centroid $00^h25^m28^s.27, + 64^{\circ}09^{'}18^{''}$ (J2000). The centroid was obtained by fitting an acceptance-corrected uncorrelated excess map with an asymmetric two dimensional Gaussian function. We used an integration radius of $0.1^{\circ}$ for the spectrum analysis. Figure~\ref{fig_tycho_spectrum_update} shows the current result along with the previous measurement. Generally both the centroid measurement and the spectrum measurement are in good agreement with published data. In this study, we extended the spectrum measurement down to $\sim$ 800 GeV. High energy events are still under systematic checking, and not included in the result.

With the entire archival data set of VERITAS, we expect to expand high energy gamma-ray measurements from several hundreds of GeV up to 10 TeV or higher and to test the existing models shown in figure~\ref{fig_tycho_spectrum}.

\vspace*{0.5cm}
\footnotesize{{\bf Acknowledgment:}{This research is supported by grants from the U.S. Department of Energy Office of Science, the U.S. National Science Foundation and the Smithsonian Institution, by NSERC in Canada, by Science Foundation Ireland (SFI 10/RFP/AST2748) and by STFC in the U.K. We acknowledge the excellent work of the technical support staff at the Fred Lawrence Whipple Observatory and at the collaborating institutions in the construction and operation of the instrument.}}

\end{document}